\documentclass{article}
\usepackage{ijcai20}

\usepackage{times}

\usepackage{soul}
\usepackage{url}
\usepackage[hidelinks]{hyperref}
\usepackage[utf8]{inputenc}
\usepackage[small]{caption}
\usepackage{graphicx}
\usepackage{amsmath}
\usepackage{booktabs}
\usepackage{amsthm,amsmath,amssymb}
\usepackage{mathrsfs}
\usepackage{multirow}
\usepackage{multicol}
\usepackage{enumitem}
\usepackage{float}

\usepackage{booktabs,caption}
\usepackage[flushleft]{threeparttable}

\usepackage{times}

\usepackage{soul}
\usepackage{url}
\usepackage[hidelinks]{hyperref}
\usepackage[utf8]{inputenc}
\usepackage{enumitem}
\usepackage{graphicx}
\usepackage{amsmath}
\usepackage{booktabs}
\usepackage{color}
\usepackage{subfigure}
\usepackage{multirow}
\usepackage{makecell}

\title{Graph Learning based Recommender Systems: A Review}

\author{
	Shoujin Wang$^1$\and
	Liang Hu$^{2,3}$\and
	Yan Wang$^2$\footnote{Corresponding author}\and
	Xiangnan He$^4$\and
	Quan Z. Sheng$^1$\and\\
	Mehmet A. Orgun$^1$\and
	Longbing Cao$^5$\and
	Francesco Ricci$^6$\and
	Philip S. Yu$^7$
	\affiliations
	$^1$ Macquarie University,
	$^2$ DeepBlue Academy of Sciences,
	$^3$ Tongji University\\
	$^4$University of Science and Technology of China,
	$^5$University of Technology Sydney\\
	$^6$Free University of Bozen-Bolzano,
	$^7$University of Illinons at Chicago\\
	\emails
	\{shoujin.wang, yan.wang, michael.sheng, mehmet.orgun\}@mq.edu.au\\
	\ longbing.cao@uts.edu.au, liandefu@ustc.edu.cn, rainmilk@gmail.com
}

\begin{document}
		
    \maketitle

    \begin{abstract}
	
Recent years have witnessed the fast development of the emerging topic of Graph Learning based Recommender Systems (GLRS). GLRS employ advanced graph learning approaches to model users' preferences and intentions as well as items' characteristics for recommendations. Differently from other RS approaches, including content-based filtering and collaborative filtering, GLRS are built on graphs where the important objects, e.g., users, items, and attributes, are either explicitly or implicitly connected. With the rapid development of graph learning techniques, exploring and exploiting homogeneous or heterogeneous relations in graphs are a promising direction for building more effective RS. In this paper, we provide a systematic review of GLRS, by discussing how they extract important knowledge from graph-based representations to improve the accuracy, reliability and explainability of the recommendations. First, we characterize and formalize GLRS, and then summarize and categorize the key challenges and main progress in this novel research area. Finally, we share some new research directions in this vibrant area.

\end{abstract}

\section{Introduction}
Recommender Systems (RS) are one of the most popular and important applications of Artificial Intelligence (AI). They have been widely 
adopted to help the users of many popular content sharing and e-Commerce web sites to more easily find relevant content, products or services. Meanwhile, Graph Learning (GL), which relates to machine learning applied to graph structure data, is an emerging technique of AI which is rapidly developing and has shown its great capability in recent years~\cite{wu2020comprehensive}. In fact, by benefiting from these capabilities to learn relational data, an emerging RS paradigm built on GL, i.e., Graph Learning based Recommender Systems (GLRS), has been proposed and studied extensively in the last few years~\cite{guo2020survey}. In this paper we offer a systematic review of the challenges and progresses in this emerging area.

\begin{figure}
    \centering
    \includegraphics[width=0.95\columnwidth]{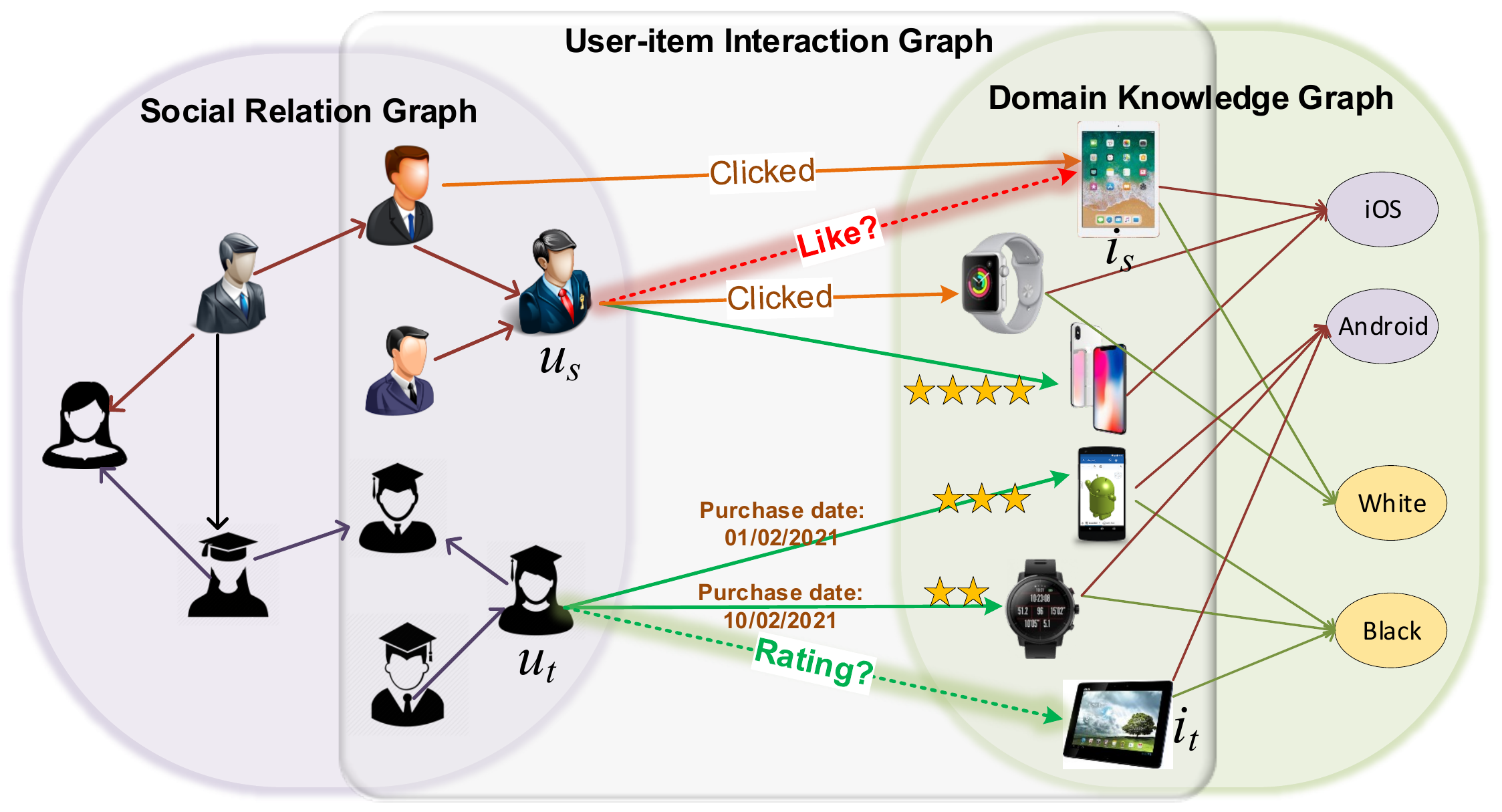}
    \vspace{-0.5em}
     \caption{The demonstration of graph learning based recommender systems }\label{fig:toy}
    \vspace{-1.5em}
\end{figure}

\subsubsection{Motivation: why graph learning for RS?}

\paragraph{Most of the data in RS has essentially a graph structure.} In the real world, most of the objects around us are explicitly or implicitly connected with each other; in other words, we are living in a world of graphs. Such characteristic is even more obvious in RS where the objects here considered including users, items, attributes, context, are tightly connected with each other and influence each other via various relations \cite{hu2014deep}, as shown in Figure \ref{fig:toy}. 
In practice, various kinds of graphs arise from the data used by RS, and they can significantly contribute to the quality of the recommendations. 

\paragraph{Graph learning has the capability to learn complex relations.} As one of the most promising machine learning techniques, GL has shown great potential in deriving knowledge embedded in different kinds of graphs. Specifically, many GL techniques, such as random walk and graph neural networks, have been developed to learn the particular type of relations modeled by graphs, and have demonstrated to be quite effective~\cite{wu2020comprehensive}. Consequently, employing GL to model various relations in RS is a natural and compelling choice.

\subsubsection{Formalization: how does graph learning can help RS?}
To date, there is no unified formalization of GLRS. 
We generally formalize GLRS from a high-level perspective. 

We construct a graph $\mathcal{G}=\{\mathcal{V},\mathcal{E}\}$ with the data of an RS where the objects, e.g., users and items, are represented as nodes in $\mathcal{V}$ and the relations between them, e.g., purchases, are represented as edges in $\mathcal{E}$. Then, a GLRS model $M(\Theta)$ is constructed and trained to generate optimal recommendation results $\mathcal{R}$ with optimized model parameters $\Theta$ that are learned from the topological and content information of $\mathcal{G}$. Formally,      

\begin {equation}
\label{eq:GLRS}
  \mathcal{R} = \arg \max_\Theta f(M(\Theta)|\mathcal{G}).
\end {equation} 

Depending on the specific recommendation data and scenarios, the graph $\mathcal{G}$ and the recommendation target $\mathcal{R}$ can be defined in various forms, e.g., $\mathcal{G}$ can be homogeneous sequences or heterogeneous networks while $\mathcal{R}$ can be predicted ratings or ranking over items. The objective function $f$ can be the maximum utility~\cite{wang2019kgat} 
or the maximum probability to form links between nodes~\cite{verma2019heterogeneous}.

\textbf{Contributions}.
The main contributions of this work are summarized below:
\begin{itemize}[leftmargin=*]
\item We systematically analyze the key challenges presented by various GLRS graphs and categorize them from a data driven perspective, providing a useful view to better understand the important characteristics of GLRS.

\item We summarize the current research progress in GLRS by systematically categorizing the more technical state-of-the-art literature. 

\item We share and discuss some open research directions of GLRS for giving references to the community.

\end{itemize}

\section{Data Characteristics and Challenges}\label{Challenges}

Different objects are managed by an RS, e.g., users, items, attributes. All of them are inter-connected with various types of relations \cite{hu2019hers}, e.g., social relations between users, or interactions between users and items. This results in different types of graphs that may be considered in an RS. 
In this section, we first classify the various types of data used in RS by considering their source and characteristics. For each class, we analyze its characteristics, then discuss how to better represent it with graphs, and finally indicate challenges that these characteristics pose to building GLRS. A brief summary of these data types is provided in Table \ref{tab:Challenge}.

It is well known that the three key objects managed by an RS are \textit{user}, \textit{item} and \textit{user-item interaction} (interaction for short), and thus all the data managed by an RS is related to them. There are two broad types of data: \textit{user-item interaction data}, e.g., clicks and ratings of users for items, and \textit{side information data}, e.g., users' profiles and items' attributes \cite{shi2014collaborative}. Depending on whether the temporal order of the interactions is recorded or not, interaction data can be classified into sequential interaction data and general interaction data. Hence, we classify the data of an RS into three classes: (1) \textit{general interaction data}, (2) \textit{sequential interaction data}, and (3) \textit{side information data}. Each class can be further divided into multiple sub-classes (cf. Table \ref{tab:Challenge}).

\begin{table*}[htbp]
  \centering
  \caption{A summary of data in RS, the representing graph, and the corresponding GLRS approach}
  \vspace{-0.5em}
  \scalebox{.75}{
    \begin{tabular}{l|l|l|l}
    \toprule
    \multicolumn{1}{c|}{Data Class} & \multicolumn{1}{c|}{Data Subclass} & \multicolumn{1}{c|}{Representing Graph} 
     & \multicolumn{1}{c}{Representative Approach Category} \\
    \midrule
    \multirow{2}[4]{*}{General interaction} & Explicit interaction & Weighted bipartite graph & Graph auto-encoder$^{[1]}$, GCN$^{[2]}$, GraphSage$^{[3]}$  \\
\cmidrule{2-4}   & Implicit interaction & Unweighted bipartite graph & Random walk$^{[4]}$, graph embedding$^{[5]}$, GCN$^{[6]}$, GraphSage$^{[7]}$  \\
    \midrule
    \multirow{2}[4]{*}{Sequential interaction} & Single-type interactions & Directed homogeneous graph &  GGNN$^{[8]}$, GraphSage$^{[9]}$, GAT$^{[10]}$  \\
\cmidrule{2-4}   & Multi-type interactions & Directed heterogeneous graph & GraphSage$^{[11]}$ \\
    \midrule
    \multirow{3}[6]{*}{Side information} & Attribute information & Heterogeneous graph & Graph embedding$^{[12]}$, GAT$^{[13]}$  \\
    \cmidrule{2-4}   & Social information &  Homogeneous graph & Random walk$^{[14]}$, graph embedding$^{[15]}$, GAT$^{[16]}$ \\
    \cmidrule{2-4}  & External knowledge & Tree or heterogeneous graph & Graph embedding$^{[17]}$, GCN$^{[18]}$  \\
  \midrule
    \multicolumn{4}{l}{\scalebox{.76}{ $^{[1]}$\cite{berg2017graph};$^{[2]}$\cite{monti2017geometric};$^{[3]}$\cite{zhang2019inductive};$^{[4]}$\cite{nikolakopoulos2019recwalk};$^{[5]}$\cite{chen2019collaborative};$^{[6]}$\cite{he2020lightgcn};$^{[7]}$\cite{zheng2018spectral};$^{[8]}$\cite{wu2019session};$^{[9]}$\cite{ma2020memory} } }\\
    
     \multicolumn{4}{l}{\scalebox{.76}{$^{[10]}$\cite{qiu2019rethinking};$^{[11]}$\cite{wang2020beyond};$^{[12]}$\cite{shi2018heterogeneous};$^{[13]}$\cite{wang2019kgat};$^{[14]}$\cite{jamali2009trustwalker};$^{[15]}$\cite{wen2018network};$^{[16]}$\cite{fan2019graph};$^{[17]}$\cite{gao2019explainable};$^{[18]}$\cite{wang2019knowledge}}   } \\
    \bottomrule
    \end{tabular}%
    }
  \label{tab:Challenge}%
  \vspace{-1.2em}
\end{table*}%

\subsection{GLRS Built on General Interaction Data}\label{General}
Interactions between users and items are usually represented as an interaction matrix, where each row indicates one user and each column indicates one item. Each entry in the matrix captures an information about the type of occurred interaction.
Depending on the interaction type, the interaction data can be divided into the explicit (i.e., users' ratings on items) and the implicit (e.g., click, view)~\cite{zhang2019deep}. Then, the recommendation task based on this general interaction data is usually formulated as a matrix completion task~\cite{zhang2019inductive}.

An interaction matrix can be naturally represented as a \textit{user-item bipartite graph}~\cite{zha2001bipartite}. In this graph, the user nodes and the item nodes constitute the two ``parts'' respectively, while the interactions are represented as edges connecting two nodes in different parts. Furthermore, an explicit interaction matrix can be represented as a \textit{weighted bipartite graph} where each edge is labeled with a weight to indicate the rating value. An implicit interaction matrix can be represented as an \textit{unweighted bipartite graph} where en edge indicates a implicit interaction. Hence, from this graph based perspective, the recommendation task is converted to link prediction on the RS bipartite graph~\cite{li2013recommendation}.

The advantage of building a GLRS on a bipartite graph is obvious. Since most users often interacted with only a small proportion of the large amount of items, 
matrix completion methods generally face data sparsity related and cold-start problems as discussed in~\cite{jamali2009trustwalker}. A bipartite graph based approach mitigates these issues by enabling the information propagating widely among nodes to enrich the information of those users and items with less interactions~\cite{wu2020graph}. However, it is challenging \textit{how to effectively and efficiently propagate the information between users or items.} This is particularly challenging in a bipartite graph since no direct links exist between users or items, and thus the information should be propagated via multi-hop neighbour nodes. For instance, to propagate some information from user $u_1$ to a similar user $u_2$ one needs to first propagate it to a bridge item $v_1$ connecting both users, and then to $u_2$ from $v_1$.

While targeting this challenge, a variety of GLRS approaches have been developed. For weighted bipartite graphs, there are mainly graph auto-encoder (e.g., graph convolutional matrix completion~\cite{berg2017graph}), Graph Convolutional Networks (GCN) (e.g., multi-graph convolutional neural networks~\cite{monti2017geometric}, stacked and reconstructed GCN~\cite{zhang2019star}), and Graph SAmple and aggreGatE (GraphSage) (e.g., inductive graph-based
matrix completion~\cite{zhang2019inductive}). For unweighted bipartite graphs, there are mainly random walk (e.g., RecWalk~\cite{nikolakopoulos2019recwalk}), graph embedding (e.g., high-order proximity for implicit recommendation~\cite{yang2018hop}, collaborative similarity embedding~\cite{chen2019collaborative}), GCN (e.g., spectral collaborative filtering~\cite{zheng2018spectral}, lightGCN~\cite{he2020lightgcn}, low-pass collaborative filter~\cite{yu2020graph}, multi-behavior GCN~\cite{jin2020multi}), and GraphSage (e.g., neural graph collaborative filtering~\cite{zheng2018spectral}). The gist of these approaches will be discussed in Section~\ref{Approaches}.

\subsection{GLRS Built on Sequential Interaction Data}\label{Sequential}
A sequential interaction data set is a collection of sequences of user-item interactions (e.g., click, purchase) registered during a given time period, and ordered by their time stamp. According to the number of interaction types included in a sequence, a sequential interaction data set can be divided into \textit{single-type interaction data set} where only one type of interactions is included, and \textit{multi-type interaction data set} where multiple types of interactions are included. Multi-type interactions like view, click and purchase co-happening in one sequence are very common in practice~\cite{wang2020beyond}. For a given user $u$, a single-type interaction sequence is usually recorded as a sequence of interacted with items (denoted as $v$), e.g., $\{v_1,...,v_n\}$, while a multi-type interaction sequence is recorded as a sequence of $\langle interaction \ type, item \rangle$ pairs, e.g., $\{click \ v_1, click \ v_2,...,purchase \ v_n\}$. An RS built on sequential interaction data is formalized as a Sequential Recommender System (SRS) which takes a sequence of historical interactions as input to predict the possible next interaction(s) 
~\cite{quadrana2018sequence,wang2019sequential}.    

A sequential interaction data set can be represented as a directed graph where each interaction sequence corresponds to one path in the graph~\cite{wu2019session}. In each path, the interactions serve as the nodes and a directed edge between any adjacent nodes indicates the order of interactions. In a multi-type interaction sequence, each element is a $\langle interaction \ type, item \rangle$ pair, which results in a compound node composed of two parts. Note that in some cases one user may have multiple identical interactions happening in a sequence (e.g., click the same items multiple times), resulting in a path consisting of one or more loops~\cite{wu2019session}. 

The advantages of building SRS on directed graph lies in the strong capability of graph learning to represent and model even the most complicated transitions in a sequence of interactions. There are usually complicated transitions which deviate from simple one-way consecutive time series patterns~\cite{wang2020modelling} over sequential interactions, especially when there are multiple identical interactions in one sequence~\cite{wu2019session}. Such transitions can be well represented by the multi-direction connections in a graph and well learned by the information aggregation from neighbour nodes of different directions in graph learning~\cite{wu2019session,xu2019graph}. However, building SRS on a directed graph is still challenging. In particular it is critical \textit{how to construct a graph to effectively represent the sequential interaction data with minimal information loss, and how to propagate information on the graph to effectively model even the most complicated transitions.}    

While targeting these challenges, various SRS have been built based on graph learning. Most of the studied approaches focus on single-type interaction data, including Gated Graph Neural Networks (GGNN) (e.g., session-based recommendation with GNN~\cite{wu2019session} and graph contextualized self-attention networks~\cite{xu2019graph}), GraphSage (e.g., memory augmented GNN~\cite{ma2020memory}), and Graph ATtention networks (GAT) (e.g., full graph neural network~\cite{qiu2019rethinking}). Limited approaches for multi-type interaction data include GraphSage (e.g., multi-relational GNN for session-based prediction~\cite{wang2020beyond}).

\subsection{GLRS Incorporating Side Information Data}
Interaction data is often sparse~\cite{hu2019hers}, thus is not sufficient for correctly capturing the users' preferences and item characteristics. 
Hence, various types of side information, e.g., attribute information and social information, have been used to alleviate such an issue. In this section, we discuss three main types of side information: (1) \textit{attribute information}, (2) \textit{social information}, and (3) \textit{external knowledge}. 

\subsubsection{GLRS Incorporating Attribute Information}
Attribute information mainly includes user attributes (e.g., gender, age), and item attributes (e.g., category, price)~\cite{wang2017perceiving,han2018aspect}. 
A user (item) attribute data set is usually recorded as a user (item) information table where each row indicates one user (item) and each column is one attribute. Attribute information is often combined with general or sequential interaction data to perform recommendations.       
Given a data set, the combination of interaction data and attribute data naturally results in a \textit{heterogeneous graph}. In such a graph, three types of nodes, i.e., user node, item node and attribute value node, and at least two types of edges exist. Specifically, in the combination of general interaction data and attribute data, in addition to user-item edges (cf. Sec. ~\ref{General}), there are user (or item)-attribute value edge representing the relations between user (or item) and attributes.
In the combination of sequential interaction data and attribute data, in addition to the directed interaction-interaction edges (cf. Sec.~\ref{Sequential}), there are also item-attribute value edges. Consequently, the recommendation task here becomes the prediction of the interactions by learning the complex relations embedded in the above mentioned heterogeneous graph.

Heterogeneous graphs combine two different types of information, i.e., interaction information and attribute information, hence enabling information propagation among different types of nodes, and better coping with the mentioned data sparsity problem. However, it is challenging {\it to selectively aggregate those useful attribute information 
to improve the recommendation performance.} 

GLRS targeting such a challenge include (heterogeneous) graph embedding (e.g., entity2rec~\cite{palumbo2017entity2rec} based on node2vec, heterogeneous preference embedding~\cite{chen2016query} and heterogeneous network embedding for recommendation~\cite{shi2018heterogeneous}), and GAT (e.g., knowledge graph attention network~\cite{wang2019kgat}).

\subsubsection{GLRS Incorporating Social Information}
Social information relates to the commonly existing social relations between users. 
A particular type of social relation among the users in a data set naturally forms a homogeneous social graph where each user corresponds to a node and each social link (e.g., friend relation) between two users corresponds to an edge. 
In an RS, the social graph can be mainly used for two tasks: (1) {\it social recommendation} (recommending items to users by incorporating social information)~\cite{fan2019graph}, and (2) {\it friend recommendation} (recommending users to a given user by predicting the possible social links)~\cite{huang2015social}.         

\paragraph{Social recommendation.} Social relations enable social influence diffusion among users~\cite{wu2020diffnet++} and thus help better understand users' preferences. The combination of social information and general or sequential user-item interaction data naturally results in a heterogeneous graph comprising two parts. The first is the bipartite graph derived from the general interaction data (cf. Sec.~\ref{General}) or the directed graph extracted from the sequential interaction data (cf. Sec.~\ref{Sequential}), while the second part is the social graph connecting the users. Obviously, two heterogeneous types of information (i.e., interaction information and social information) are contained in the graph. Hence, the RS must be able to effectively leverage this heterogeneous graph to predict the unknown user-item interactions. 

Such an approach helps better understand a user's preference by considering the influence of her neighbours in a social graph. However, on one hand, it is not clear how many orders of neighbours should be considered to correctly compute this influence on a given user. On the other hand, different neighbours usually influence a user to different degrees~\cite{wu2020graph}. Hence, it is a challenge {\it to appropriately model the influence of other users to a given user.} 

Typical approaches targeting this challenge include random walk (e.g., Trust-walker~\cite{jamali2009trustwalker}), graph embedding~\cite{wen2018network} and GAT (e.g., GraphRec~\cite{fan2019graph} and improved diffusion network~\cite{wu2020diffnet++}). All these works focus on combining social graph and general interactions, while limited works combine social graph and sequential interactions~\cite{song2019session}.


\paragraph{Friend recommendation.} By using the aforementioned homogeneous social graph, friend recommendation is performed as a link prediction task on such graph~\cite{yin2010linkrec}. Specifically, given a target user and the known social links on the graph, friend recommendation first infers the possible links between other users and the target user and then recommends those users with high probabilities to link with the target user to her. The main challenge lies in {\it how to appropriately model the mutual-influence between users.} According to our analysis, random walk based approaches~\cite{backstrom2011supervised,bagci2016context} are more common in order to address this challenge. Other approaches include graph embedding~\cite{verma2019heterogeneous}. 

\subsubsection{GLRS Incorporating External Knowledge}
External knowledge, e.g., item taxonomy and semantic relations between concepts, related to users and items usually contributes to a deeper understanding of the users' preference and item characteristics~\cite{wang2018ripplenet}, and ultimately improving recommendation performance. Such knowledge is usually represented as a knowledge graph where various types of objects (e.g., users, movies, movie directors) are represented as nodes and the relations between them (e.g., movie-director relation) are represented as edges~\cite{wang2019kgat}. This graph is often combined with the graph composed by the general or sequential interaction data, giving rise to a more complex and heterogeneous graph. There are mainly two types of external knowledge commonly utilized in RS: item/user ontology and common knowledge. 

\paragraph{GLRS incorporating ontology knowledge.} The ontology of users or items is usually represented as a hierarchical tree-like graph where the hierarchical relations between users or items are recorded. A type of commonly utilized ontology knowledge for recommendations is item taxonomy information~\cite{huang2019taxonomy}. An example of such a tree graph is used in Amazon.com, where the category information of products is used to organize all the items offered by the platform. In that graph, the root node corresponds to the coarsest-grained category and the leaf nodes represent specific items.

The incorporation of item ontology knowledge enables a better understanding of the users' multi-level preferences towards items, and thus helps improving the explainability of the recommendations~\cite{gao2019explainable}. However, it remains a challenge {\it to propagate users' preferences over items along the hierarchy tree graph to extract the multi-level preferences.}     

Representative works targeting such a challenge include graph embedding based approaches~\cite{wang2018tem,gao2019explainable}, aimed at learning more informative item embedding for general recommendations, and memory network on graphs to learn coarse-grained-preference representation for sequential recommendations~\cite{huang2019taxonomy}.

\paragraph{GLRS incorporating common knowledge.} Common knowledge refers to the wide range of relations between the various entities managed by an RS. It includes, but is not limited to, general semantic relations between entities (e.g., the relations among bread, food, bakery item from Microsoft Concept
Graph\footnote{https://concept.research.microsoft.com/})~\cite{sheu2020context}, and domain-specific relations between entities (e.g., the relations between movies, directors, genre)~\cite{gao2020deep}. Due to the diversity of these entities and their relations, common knowledge is usually represented as a heterogeneous and complex graph where different types of nodes and edges exist~\cite{guo2020survey}.        

The incorporation of common knowledge benefits the exploration and exploitation of various external implicit relations between users and/or items, improving recommendation performance. However, it remains a challenge {\it to effectively propagate information between different types of entities via different types of links between them}, to obtain coherent and useful information for the recommendations.     

Representative works targeting this challenge include graph embedding methods~\cite{wang2019multi} (especially meta-path based embedding~\cite{zhao2017meta,sun2018recurrent,shi2018heterogeneous,wang2019explainable}) to wisely learn the embedding of heterogeneous entities and relations, and GNN based methods (especially GCN~\cite{wang2019knowledge} an GAT~\cite{wang2019kgat}) to iteratively aggregate the information from neighbour nodes.

\section{Graph Learning Approaches for RS}\label{Approaches}
In this section, we introduce graph learning based techniques, which offer solutions to the challenges faced by GLRS, which were discussed in Section~\ref{Challenges}. We first provide a technical categorization of the solutions, and then we discuss the gist of each solution together with the achieved progresses.

\begin{figure}
    \centering
    \includegraphics[width=0.9\columnwidth]{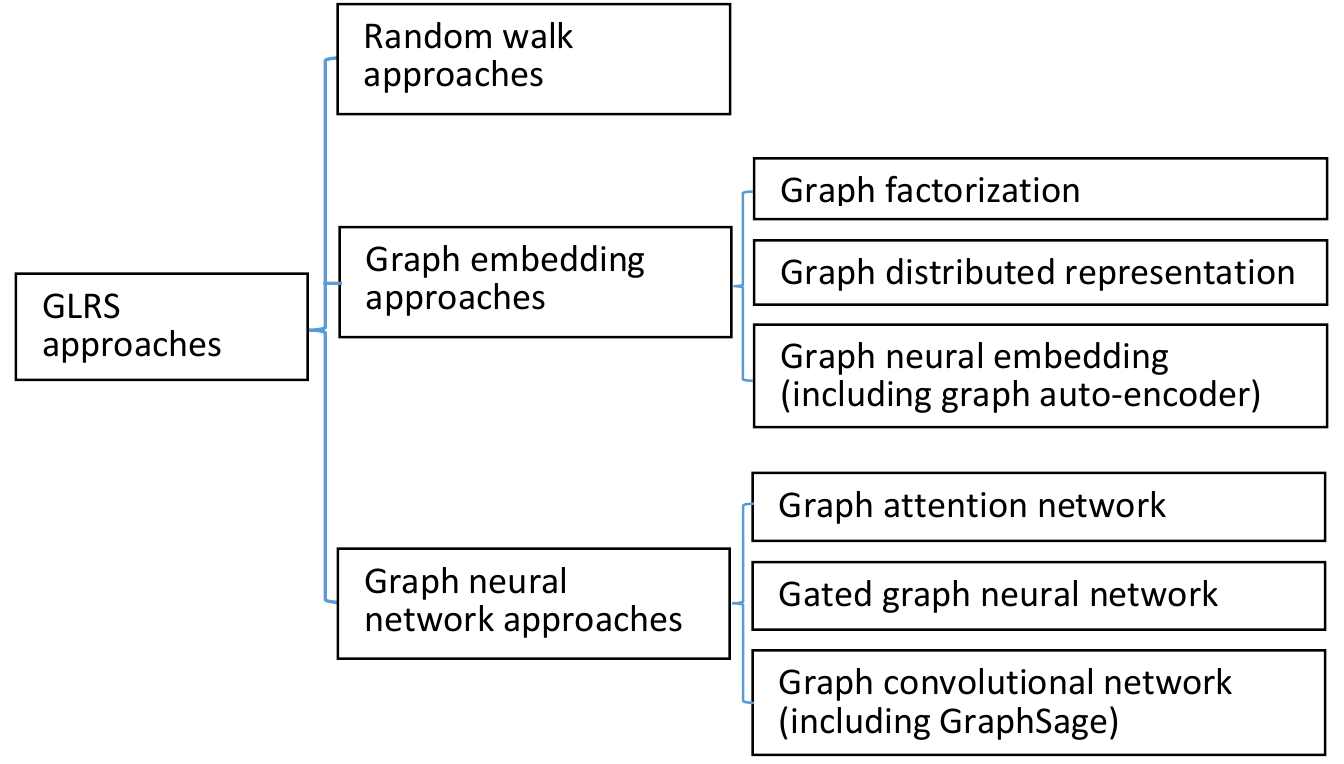}
    \vspace{-1em}
     \caption{ Classifying GLRS approaches from technical perspective}\label{fig:taxnomy}
    \vspace{-1.3em}
\end{figure}

The categorization of the approaches to GLRS is presented in Figure \ref{fig:taxnomy}. GLRS are divided into three categories, and some categories are further divided into sub-categories. 

\subsection{Random Walk Approach}\label{Progress1}
Random walk based RS have been extensively studied in the past years and have been widely employed on various types of graphs (e.g., social graphs, sequence graphs).
Generally, a random walk based RS first let a random walker to walk on a given graph with a predefined transition probability for each step, in order to model the implicit preference or interaction propagation among users and/or items, and then takes the probability the walker lands on nodes after certain steps to rank these candidate nodes for recommendations. Random walk based RS are particularly suitable for capturing the complex, higher-order and indirect relations among various types of nodes (e.g., users and items) on the graph, and thus, can address important challenges for GLRS especially those built on heterogeneous graphs. 

There are different variants of random walk based RS. Besides the basic random walk based RS \cite{baluja2008video}, random walk with restart based RS \cite{bagci2016context,jiang2018recommendation} is a representative type of several variants. It sets a constant probability to jump back to the starting node in each transition and it is generally used in graphs containing many nodes to avoid leaving the particular context of the starting node.

Although widely applied, the drawbacks of random walk based RS are clear: (1) they need to generate ranking scores on all candidate items at each step for each user, leading to low efficiency; 
(2) unlike most of the learning-based paradigms, they are heuristic-based, lacking model parameters to optimize the recommendation objective.

\subsection{Graph Embedding Approach}\label{Progress2} 
Graph embedding is an effective technique to analyze the complex relations embedded on graphs and has been rapidly developing in recent years. It maps each node into a low-dimension embedding vector which encodes the graph structure information. Researchers introduced graph embedding to model the complex relations between various nodes (e.g., users, items) and they came  up with the novel approach of Graph Embedding based RS (GERS). Depending to the specific embedding approach that is used, GERS can be divided into three classes: (1) Graph Factorization based RS (GFRS), (2) Graph Distributed Representation based RS (GDRRS), and (3) Graph Neural Embedding based RS (GNERS). 

\paragraph{Graph Factorization based RS (GFRS).} GFRS first factorizes the inter-node commuting 
matrix based on meta-path on the graph in order to obtain the embedding of each node (e.g., a user or an item), which are then used as input of the subsequent recommendation task \cite{wang2019unified}. By doing so, the complex relations between nodes in the graph are encoded into the embedding to improve the recommendations. Due to their capability to handle the heterogeneity of the nodes, GFRS have been widely applied to capture relations between different types of nodes, e.g., users and items. However, although being simple and effective, such models may easily suffer from the sparsity of the observed data.

\paragraph{Graph Distributed Representation based RS (GDRRS).}
Differently from GFRS, GDRRS usually follow Skip-gram model \cite{mikolov2013distributed} in order to learn a distributed representation of each user or item in a graph. They encode information about the user or item and its adjacent relations into a low-dimensional vector \cite{shi2018heterogeneous}, which is then used for the subsequent recommendation step. Specifically, GDRRS usually first use random walk to generate a sequence of nodes that co-occurred in one meta-path and then employ the skip-gram or similar models to generate node representations for recommendations. By exploiting its powerful capability to encode the inter-node connections in a graph, GDRRS are widely applied to both homogeneous and heterogeneous graphs for capturing the relations between the objects managed by the RS \cite{cen2019representation}. 
GDRRS have shown their great potential in recent years due to their simplicity, efficiency and efficacy.

\paragraph{Graph Neural Embedding based RS (GNERS).} GNERS utilize neural networks, like  Multilayer-perceptron, auto-encoder, to learn users or items embedding. Neural embedding models are easy to integrated with other downstream neural recommendation models (e.g., RNN based ones) to build an end-to-end RS~\cite{han2018aspect}. 
To this end, GNERS have been widely applied to a variety of graphs like attributed graphs \cite{han2018aspect}, interaction combined with knowledge graphs \cite{hu2018leveraging,cen2019representation}.

\subsection{Graph Neural Network Approach}
\label{Progress3}

Graph Neural Networks (GNN) apply neural networks techniques on graph data. 
Leveraging the strength of GNN in learning informative representations,
several RS have used GNN to address the most important challenges posed by GLRS. By considering a model perspective, GNN based RS can be mainly categorized into three classes:
(1) Graph ATtention network based RS (GATRS),
(2) Gated Graph Neural Network based RS (GGNNRS), and (3) Graph Convolutional Network (including GraphSage) based RS (GCNRS).

\begin{table*}[htb]
  \centering
  \caption{A list of representative open-source GLRS algorithms}
  \vspace{-0.5em} 
\scalebox{.75}{
    \begin{tabular}{l|l|l|l|l|l}
    \toprule
    \multicolumn{1}{c|}{Algorithm} & \multicolumn{1}{c|}{Input Data} & \multicolumn{1}{c|}{Learning Task}
     & \multicolumn{1}{c|}{Learning Approach} &\multicolumn{1}{c|}{Venue} & \multicolumn{1}{c}{Link} \\
\midrule

GC-MC$^{[1]}$  & Explicit interaction & Rating prediction & Graph auto-encoder & KDD’2018 DL &\scriptsize{\url{https://github.com/riannevdberg/gc-mc}}  \\ \cmidrule{1-6}

MGCNN$^{[2]}$ & Explicit interaction & Rating prediction &GCN & NIPS'2017 &\scriptsize{\url{https://github.com/fmonti/mgcnn}} \\ \cmidrule{1-6} 
 
IGMC$^{[3]}$ &Explicit interaction & Rating prediction & GraphSage &ICLR'2020 & \scriptsize{\url{https://github.com/muhanzhang/IGMC}} \\\cmidrule{1-6} 

RecWalk$^{[4]}$ & Implicit interaction & Click prediction& Random walk & WSDM '2019 & \scriptsize{\url{https://github.com/nikolakopoulos/RecWalk}}\\\cmidrule{1-6}

PinSage$^{[5]}$ & Implicit interaction & Click prediction & GraphSage &KDD'2018
& \scriptsize{\url{https://github.com/yoonjong12/pinsage}}\\\cmidrule{1-6} 

CSE$^{[6]}$ & Implicit interaction & Click prediction & Graph embedding & WWW'2019 & \scriptsize{\url{https://github.com/cnclabs/proNet-core}}\\\cmidrule{1-6} 
  
LightGCN$^{[7]}$ & Implicit interaction & Click prediction & GCN & SIGIR'2020 & \scriptsize{\url{https://github.com/kuandeng/LightGCN}}\\\cmidrule{1-6}   

SpectralCF$^{[8]}$ & Implicit interaction & Click prediction & GraphSage & RecSys'2018 & \scriptsize{\url{https://github.com/lzheng21/SpectralCF}}\\\cmidrule{1-6}  

SR-GNN$^{[9]}$ & Single-type sequential interaction & Next-item prediction & GGNN & AAAI'2019 & \scriptsize{\url{https://github.com/CRIPAC-DIG/SR-GNN}}\\\cmidrule{1-6}   

MA-GNN$^{[10]}$ & Single-type sequential interaction & Next-item prediction & GraphSage & AAAI'2020 & \scriptsize{\url{https://github.com/cynricfu/MAGNN}}\\\cmidrule{1-6} 
    
FGNN$^{[11]}$ & Single-type sequential interaction & Next-item prediction & GAT & CIKM'2019 & \scriptsize{\url{https://github.com/RuihongQiu/FGNN}}\\\cmidrule{1-6}

MGNN-SPred$^{[12]}$ & Multi-type sequential interaction & Next-item prediction & GraphSage & WWW'2020 & \scriptsize{\url{https://github.com/Autumn945/MGNN-SPred}}\\\cmidrule{1-6}

HERec$^{[13]}$ & Explicit interaction + Attribute information & Rating prediction & Graph embedding & TKDE'2018 & \scriptsize{\url{https://github.com/librahu/HERec}}\\\cmidrule{1-6}

KGAT$^{[14]}$ & Implicit interaction + Attribute information & Click prediction & GAT & KDD'2019 & \scriptsize{\url{https://github.com/xiangwang1223/}}\\\cmidrule{1-6}

TrustWalker$^{[15]}$ & Explicit interaction + Trust relation & Rating prediction & Random walk & KDD'2009 & \scriptsize{\url{https://github.com/Antili/TrustWalker}}\\\cmidrule{1-6}

GraphRec$^{[16]}$ & Explicit interaction + Social relation & Rating prediction & GAT & WWW'2019 & \scriptsize{\url{https://github.com/wenqifan03/GraphRec-WWW19}}\\\cmidrule{1-6}

KGNN-LS$^{[17]}$ & Implicit interaction + External knowledge & Click prediction & GAT & KDD'2019 & \scriptsize{\url{https://github.com/hwwang55/KGNN-LS}}\\\cmidrule{1-6}

\multicolumn{6}{l}{\scalebox{.76}{ $^{[1]}$ \cite{berg2017graph};$^{[2]}$\cite{monti2017geometric};$^{[3]}$\cite{zhang2019inductive};$^{[4]}$\cite{nikolakopoulos2019recwalk};$^{[5]}$\cite{ying2018graph};

$^{[6]}$\cite{chen2019collaborative};$^{[7]}$\cite{he2020lightgcn};$^{[8]}$\cite{zheng2018spectral};$^{[9]}$\cite{wu2019session}; } }\\
    
\multicolumn{6}{l}{\scalebox{.76}{$^{[10]}$\cite{ma2020memory};$^{[11]}$\cite{qiu2019rethinking};$^{[12]}$\cite{wang2020beyond};$^{[13]}$\cite{shi2018heterogeneous};$^{[14]}$\cite{wang2019kgat};$^{[15]}$\cite{jamali2009trustwalker};$^{[16]}$\cite{fan2019graph};$^{[17]}$\cite{wang2019knowledge}} } \\
    \bottomrule
    \end{tabular}%
    }

   \begin{tablenotes}
    \item[2]
    {
    \scriptsize 2. More up-to-date resource on GLRS can be found at: \url{https://github.com/shoujin88/graph-learning-based-recommender-systems-GLRS-}}.
    \end{tablenotes} 
    
  \label{tab:code}%
\end{table*}%

\paragraph{Graph ATtention network based RS (GATRS).}
Graph ATtention networks (GAT) introduce attention mechanisms into GNN to discriminatively learn the different relevance and influence degree of other users (items) w.r.t. the target user (item) on a given graph. 
GATRS are based on GAT for precisely learning inter-user or item relations. In such a case, the influence of the more important users or items, w.r.t. a specific user or item, is emphasized, which is more in line with the real-world cases and this has been shown to be beneficial for the recommendations. 
Due to their good discrimination capability, GAT are widely used in different kinds of graphs including social graphs \cite{fan2019graph}, item session graphs \cite{xu2019graph}, and knowledge graphs \cite{wang2019kgat}.

\paragraph{Gated Graph Neural Network based RS (GGNNRS).}
Gated graph neural networks (GGNN) introduce the Gated Recurrent Unit (GRU) into GNN to learn the optimized node representations by iteratively absorbing the influence of other nodes in a graph to comprehensively capture the inter-node relations. GGNNRS are built on GGNN to learn the user or item embeddings for recommendations by comprehensively considering the complex inter-user or inter-item relations. 
Due to their capability to capture the complex relations between nodes, GGNN are widely used to model the complex transitions between items in a sequence graph for sequential recommendations \cite{wu2019session}, or to model the complex interactions between different categories of fashion products for fashion recommendations \cite{cui2019dressing}, and they have achieved superior recommendation performance.

\paragraph{Graph Convolutional Network based RS (GCNRS).}
Graph Convolutional Networks (GCN) generally learn how to iteratively aggregate feature information from local graph neighbor nodes by leveraging both graph structure and node feature information. In general, by utilizing the convolution and pooling operations, GCNs are capable of learning informative embeddings of users and items by effectively aggregating information from their neighborhoods in graphs. GCNRS are built on GCN to learn the user or item embeddings in a graph while exploiting both the complex relations between users or/and items and their own content information for recommendations \cite{ying2018graph}. Thanks to the powerful feature extraction and learning capability, particularly their strength in combining the graph structure and node content information, GCN are widely applied to a variety of graphs in RS to build GCNRS and are demonstrated to be very effective. For instance, GCN are used for influence diffusion on social graphs in social recommendations \cite{wu2019neural}, for mining the hidden user-item connection information on user-item interaction graphs, for alleviating the data sparsity problemn in collaborative filtering \cite{wang2019binarized}, and for capturing inter-item relatedness by mining their associated attributes on knowledge graphs \cite{wang2019knowledge}.

\section{GLRS Algorithms and Datasets}

The source code of most of the representative GLRS algorithms is publicly accessible. In Table~\ref{tab:code}, to facilitate the access for empirical analysis, we summarize source codes of algorithms for GLRS which take various input data and use different learning approaches for different learning tasks. The listed algorithms are carefully selected and are commonly used as baselines in existing work. 

In addition to algorithms, datasets are another important part for empirical analysis of GLRS approaches. In order to facilitate the empirical analysis of the surveyed algorithms, in Table~\ref{tab:data} we also list public and real-world datasets with different characteristics from various domains. These datasets are commonly used for evaluating GLRS algorithms.

\begin{table*}[!htb]
  \centering
  \caption{A list of commonly used and publicly accessible real-world datasets for GLRS}
  \vspace{-0.5em}
  
\scalebox{.75}{
    \begin{tabular}{l|l|l|l|l|l}
    \toprule
    \multicolumn{1}{c|}{Dataset} & \multicolumn{1}{c|}{Domain} & \multicolumn{1}{c|}{Information Included} & \multicolumn{1}{c|}{\# Interactions}
     & \multicolumn{1}{c|}{Reference} & \multicolumn{1}{c}{Link} \\
\midrule

MovieLens-1M & Movie & Explicit interaction & 1,000,209 & \cite{zheng2018spectral} &\scriptsize{\url{https://grouplens.org/datasets/}}  \\ \cmidrule{1-6} 

HetRec & Movie & Explicit interaction & 855,598 & \cite{zheng2018spectral} &\scriptsize{\url{https://grouplens.org/datasets/}}  \\ \cmidrule{1-6}

Amazon instant video & Video & Explicit interaction & 583,933 & \cite{zheng2018spectral} &\scriptsize{\url{http://jmcauley.ucsd.edu/data/amazon/}}  \\ \cmidrule{1-6}

Gowalla  & POI & Implicit interaction & 1,027,370 & \cite{he2020lightgcn} &\scriptsize{\url{http://snap.stanford.edu/data/loc-gowalla.html}}  \\ \cmidrule{1-6}
   
Yelp 2018 & POI & Implicit interaction&  1,561,406 & \cite{he2020lightgcn} &\scriptsize{\url{https://www.yelp.com/dataset}}  \\ \cmidrule{1-6}

Amazon-book & E-commerce & Implicit interaction &  2,984,108 & \cite{he2020lightgcn} &\scriptsize{\url{https://github.com/uchidalab/book-dataset}}  \\ \cmidrule{1-6}

Yoochoose 1/4 & E-commerce & Stream of clicks &8,326,407 & \cite{wu2019session} &\scriptsize{\url{http://2015.recsyschallenge.com/challege.html}} \\ \cmidrule{1-6} 
 
Diginetica & E-commerce & Stream of clicks &982,961 & \cite{wu2019session} &\scriptsize{\url{https://competitions.codalab.org/competitions/11161}} \\ \cmidrule{1-6}

Book-crossing & Book & Ratings and attribute information & 1,000,000 & \cite{wang2019knowledge} &\scriptsize{\url{http://www2.informatik.uni-freiburg.de/~cziegler/BX/}} \\ \cmidrule{1-6}  
 
Last.FM & Music & Implicit interaction, social and tag &  92,834 & \cite{wang2019knowledge} &\scriptsize{\url{https://grouplens.org/datasets/hetrec-2011/}} \\ \cmidrule{1-6}  
 
Epinions & E-commerce & Rating, trust relation &  764,352 & \cite{fan2019graph} &\scriptsize{\url{http://alchemy.cs.washington.edu/data/epinions/}} \\ \cmidrule{1-6}  
 
Ciao& E-commerce & Rating, trust relation &  283,319 & \cite{fan2019graph} &\scriptsize{\url{https://www.cse.msu.edu/~tangjili/datasetcode/truststudy.htm}} \\ \cmidrule{1-6}

Amazon-toys and games& E-commerce & Implicit interaction & 167,597 & \cite{gao2019explainable} &\scriptsize{\url{ http://jmcauley.ucsd.edu/data/amazon}} \\ \cmidrule{1-6}

Amazon-digital music& Music & Implicit interaction &  64,706 & \cite{gao2019explainable} &\scriptsize{\url{ http://jmcauley.ucsd.edu/data/amazon}} \\ 

    \bottomrule
    \end{tabular}%
    }
  \label{tab:data}%
  \vspace{-1.em}
\end{table*}%

\section{Open Research Directions}

GLRS are fast developing. Although substantial results have been achieved, 
some challenges still remain. By matching the demonstrated challenges to the research progress already achieved, we have identified some open research directions.

\paragraph{Self-evolutionary RS with dynamic-graph learning.} In real-world RS, users, items and the interactions between them, keep evolving over time~\cite{wang2019survey}. This originates graphs with dynamic topology, and such dynamics could have direct impacts on the user and requirement modeling, causing even a clear change of recommendation results over time. However, this issue is still underestimated in existing GLRS. Therefore, it is a promising future research direction to design self-evolutionary RS over dynamic graphs.\vspace{0.1em}

\paragraph{Explainable RS with causal graph learning.} Causal inference is a major technique used to discover the causal relations between objects or actions. Although some progress has been achieved in explainable RS, we are is still far away from achieving a complete understanding of the reasons and intents behind user choice behaviours, which is a critical step to make reliable and explainable recommendations~\cite{zhang2018explainable}. To this end, it is another promising direction to construct explainable RS with causal graph learning.          

\noindent 
\paragraph{Cross-domain RS with multiplex graph learning.} In reality, the data and interactions for recommendation could be derived from multiple domains, including various sources, systems, and modalities~\cite{zhu2019dtcdr}. These are inter-correlated and must collaboratively contribute to the recommendations~\cite{zhu2021cross}. Consequently, the interactions in cross-domain RS can be represented by multiplex networks where nodes may or may not be interconnected with other nodes in other layers. As a result, the new generation cross-domain RS potentially works with multiplex graph learning.

\noindent 
\paragraph{High-efficiency online RS with large-scale graph learning.}
An inevitable issue in real RS is the scale of data, which is often large and leads to high cost in terms of both time and space. This issue is even more important in GLRS since the graph structure data is usually even larger and requires more time and space to be processed, let alone to perform complex machine learning techniques on it to generate recommendations. Therefore, it is necessary to study more  efficient algorithms to speed up large-scale online processing and learning to keep updating model to generate 
timely recommendations.

\section{Conclusions}
As one of the most important applications of Artificial Intelligence (AI), Recommender Systems (RS) can be found nearly at every corner of our daily lives. Graph Learning (GL), as one of the most promising AI techniques, has shown a great capability to learn the complex relations among the various objects managed by an RS. This has launched  a totally new RS paradigm: Graph Learning based Recommender Systems (GLRS), which is of great potential to be the next-generation of RS. It is our hope that this review has provided a comprehensive and  self contained overview of the  recent progress, challenges as well as future research directions in GLRS to both the academia and industry.

\section*{Acknowledgements} This work was supported by ARC 
Discovery Project DP180102378.

{
\bibliographystyle{named}
\small{
\bibliography{ijcai20}}
}
\end{document}